\documentstyle[12pt]{article}

\begin{document}
\title{\textbf{Mean-field cosmic dynamo and curvature vs turbulence spectrum in Riemannian space}} \maketitle
{\sl \textbf{L.C. Garcia de Andrade}\newline
Departamento de F\'{\i}sica
Te\'orica-IF-UERJ- RJ, Brasil\\[-3mm]
\vspace{0.01cm} \paragraph*{Previous attempts for building a cosmic dynamo including preheating in inflationary universes [Bassett et al Phys Rev (2001)] has not included mean field dynamos. Here, a mean field dynamo in cosmic scales on a Riemannian spatial cosmological section background, is set up. When magnetic fields and flow velocities are parallel propagated along the Riemannian space dynamo action is obtained. Turbulent diffusivity ${\beta}$ is coupled with the Ricci magnetic curvature, as in Marklund and Clarkson [MNRAS (2005)], GR-MHD dynamo equation. Mean electric field possesses an extra term due to Ricci tensor coupling with magnetic vector potential in Ohm's law. Goedel universe induces a mean field dynamo growth rate ${\gamma}=2{\omega}^{2}{\beta}$. In this frame kinetic helicity vanishes. By considering a universe vorticity, ${\omega}\approx{10^{-16}s^{-1}}$ for galactic dynamos, thus ${\gamma}=2.10^{-32}{\beta}$, and since ${\beta}\approx{10^{26}cm^{2}s^{-1}}$, the growth rate ${\gamma}\approx{10^{-6}s^{-1}}$. In non-comoving the magnetic field is expressed as $B\approx{\sqrt{\frac{2{\beta}}{\gamma}}{\times}10^{-6}G}\approx{10^{10}G}$ a magnetic field found in the nucleosynthesis era. The Ricci scalar turbulence spectrum of the cosmic dynamos is computed from the GR-MHD dynamo equation. By analyzing the Fourier modes of the Ricci scalar, one shows that the curvature energy spectrum of the turbulent dynamo is similar to the Kolmogorov spectrum. Recently Mizeva et al [Doklady Physics (2009)] have shown the the slope of the turbulent spectrum may grow from ${-1.88}$ to $2$, as shown here from curvature spectrum in turbulent dynamos. Similar enhancements of turbulence in Friedmann cosmology have been obtained by Brandenburg et al [Phys Rev D (1997)].}{PACS: 47.65.Md, 02.40-Ky. Key-words: Dynamo plasmas; Riemannian geometry. }
\newpage
\section{Introduction}
The real astronomical universe is certainly turbulent. Either inside black holes or galactic nucleus or even in the early universe, things do not behave in the very simple realm of laboratory physics of lo velocity hydrodynamics and cosmology even if relativistic is not so laminar as in the present stages of universe expansion. On the other hand dynamo theory \cite{1} has been one of the most successful theory able to explain galactic \cite{2} and solar magnetism \cite{3}. The mean field dynamo theory developed mainly by Raedler and Krause \cite{4}, paved the way to the better understand of the randomic processes in the turbulent plasma. From the mathematical point of view, the first simple solution of dynamo equation in Riemannian space was given  by Arnold et al \cite{5}, which have made use of a somewhat unrealistic hypothesis of uniform stretching in Riemannian spaces given by steady flows. More realistic Riemannian solutions in pseudo-Anosov space has been presented by Gilbert \cite{6} following the where particles and magnetic field lines were stretched along a torus map. Another sort of fast dynamo as presented by Chicone et al \cite{2} as a two-dimensional compact stretched Riemannian manifold. Chicone et al proved that in order this manifold could host a fast dynamo, the Riemannian curvature had to be constant and negative. Yet more recently Garcia de Andrade \cite{7} has presented two solutions in three-dimensional stretched Riemannian space representing fast dynamos in plasmas, all of these solutions of dynamo Faraday equation are not very realistic because they basically do not take into consideration oscillation in the fields and fluctuations of random fields, which happens naturally in mean field theory as presented by Raedler and Krause \cite{4}. \newline
Investigation of dynamo action on those Riemannian manifolds are fundamental to obtain the connection between dynamo theory and experiment in unusual topological flows as recently shown Moebius strip dynamo flows \cite{8}, in the case of the Perm liquid sodium dynamo experiment. In this last example of slow dynamo, a dynamo wave was obtained along the twist directions of the Moebius dynamo flow. Question why to link Riemannian 2-manifold of constant negative curvature to dynamos, or a Anosov space to dynamos, is responded by the Chicone et al recently work who showed that, fast dynamos in 2D can only be realized in Riemannian spaces of constant negative curvature, without violating Cowling anti-dynamo theorem \cite{9}. In this paper, one obtains expressions for the Faraday self-induction equation in Riemannian space of arbitrary curvature, starting from a review on plasma dynamos meanfields. The result is applied to Goedel cosmology, to compute the growth rate of magnetic fields in that universe from COBE data. \newline
This paper is organised as follows: Section II addresses the review of meanfield dynamo plasmas and present the ne result of the computation of the mean electric field in Riemannian space. In section III one shows that the diffusive magnetic dynamo may be expressed in Riemannian space in terms of kinetic helicity and turbulent diffusivity. In section IV results are applied to Goedel universe mean field dynamos. In section V the Ricci scalar turbulent spectrum of the GR MHD dynamo equation is given. Future prospects are presented in section VI.
\section{Electric mean fields in Riemannian spaces}
In this section one shall be concerned with the derivation of electric mean field in the three-dimensional Riemann space. To start with let us express review briefly the induction equation in terms of the parameters of the plasma flow
\begin{equation}
\textbf{B}(\textbf{r},t)=\textbf{B}_{0}(\textbf{r},t)+\textbf{b}(\textbf{r},t)
\label{1}
\end{equation}
and
\begin{equation}
\textbf{U}(\textbf{r},t)=\textbf{U}_{0}(\textbf{r},t)+\textbf{u}(\textbf{r},t)
\label{2}
\end{equation}
where S is the Lundquist number and ${\eta}$ is the non-turbulent diffusivity parameter. Where random fields representing magnetic and flow velocity fields are respectively by
\begin{equation}
\langle\textbf{b}(\textbf{r},t)\rangle=0
\label{3}
\end{equation}
and
\begin{equation}
\langle\textbf{u}(\textbf{r},t)\rangle=0
\label{4}
\end{equation}
where $<...>$ represents the time space average of the inside brackets quantities. Note that this splitting transform the Faraday self-induction equation into the dynamo plasma equation
\begin{equation}
{\partial}_{t}\textbf{B}_{0}(\textbf{r},t)={\nabla}{\times}(\textbf{U}_{0}{\times}\textbf{B}_{0})-{\nabla}{\times}
\textbf{E}_{f}-\frac{\eta}{S}{\nabla}{\times}({\nabla}{\times}\textbf{B}_{0})
\label{5}
\end{equation}
and
\begin{equation}
\textbf{U}(\textbf{r},t)=\textbf{U}_{0}(\textbf{r},t)+\textbf{u}(\textbf{r},t)
\label{6}
\end{equation}
where $\textbf{E}_{f}$ is the mean electric field given by
\begin{equation}
\textbf{E}_{f}=-\langle\textbf{u}{\times}\textbf{b}\rangle
\label{7}
\end{equation}
where the Ohm's law is given by
\begin{equation}
\textbf{E}_{f}=-\textbf{U}{\times}\textbf{B}+{\nabla}{\phi}+\frac{\eta}{S}\textbf{J}
\label{8}
\end{equation}
This mean field dynamo is very important in turbulent dynamos. This justifies that one writes down an expression for this field in Riemannian space as an exercise for the mean field dynamos in Riemannian space in the next section. Here ${\phi}$ is the electric potential and $\textbf{J}$ is the electric current. A one shall now show, the interaction with the Riemannian curvature presents certain peculiarities and introduces certain extra terms that may be important in future analysis of turbulent cosmic dynamos such as ones that appears in magnetars for example. A simple computation of the electric mean field curl, which balances the diffusivity terms, even when the laminar contributions satisfies the symmetries of Cowling's anti-dynamo theorem. This computation yields
\begin{equation}
{\nabla}{\times}
\textbf{E}_{f}=-{\nabla}{\times}\langle{\textbf{u}{\times}\textbf{b}}\rangle
\label{9}
\end{equation}
Noting that the a simple and straightforward computation allow us to reach the following expression
\begin{equation}
[{\nabla}{\times}\textbf{E}_{f}]^{i}=-\frac{1}{\sqrt{g}}{\partial}_{j}\langle{\sqrt{g}({u}^{i}b^{j}-u^{j}b^{i})}\rangle
\label{10}
\end{equation}
Even without completing this computation, it is easy to note that the Riemannian effects shall be present in terms of the Riemann-Christoffel symbols since
\begin{equation}
{\Gamma}_{j}=\frac{1}{\sqrt{g}}{\partial}_{j}[\sqrt{g}]
\label{11}
\end{equation}
where ${\Gamma}_{j}={{\Gamma}^{i}}_{ji}$ which is the trace of the Riemann-Christoffel symbols
\begin{equation}
{{\Gamma}^{i}}_{jk}=\frac{1}{2}g^{il}[g_{lj,k}+g_{lk,j}-g_{jk,l}]
\label{12}
\end{equation}
where comma denotes the partial derivative with respect to the spatial coordinates. Therefore the Riemann-Christoffel trace shall couple with the random part of the term inside the brackets in (\ref{11}). By considering the electric potential in the gauge
\begin{equation}
{\phi}={\eta}{\nabla}.\textbf{A}
\label{13}
\end{equation}
By expanding the equation
\begin{equation}
{\nabla}.\textbf{A}=\frac{1}{\sqrt{g}}{\partial}_{i}(\sqrt{g}A^{i})={\partial}_{i}A^{i}+{\Gamma}_{i}A^{i}
\label{14}
\end{equation}
Thus by substituting this expression into the contribution of the electric potential to the electric mean field yields
\begin{equation}
{\nabla}{\phi}={\eta}{\nabla}[{\nabla}.\textbf{A}]
\label{15}
\end{equation}
yields the electric mean field as
\begin{equation}
\textbf{E}_{f}=-{\eta}\langle{Ric,\textbf{A}}\rangle+{\nabla}\textbf{J}+\textbf{U}{\times}\textbf{B}
\label{16}
\end{equation}
where $\langle{Ric,\textbf{A}}\rangle$ is given explicitly on a coordinate chart by the inner public
\begin{equation}
\langle{Ric,\textbf{A}}\rangle={R^{i}}_{j}A^{j}
\label{17}
\end{equation}
Here the Ricci tensor is given on a coordinate-free language by
\begin{equation}
Ric={R_{ij}}dx^{i}{\otimes}dx^{j}
\label{18}
\end{equation}
where $\otimes$ represents here the tensor product. Though in principle we are not here considering the pseudo-Riemannian spacetime of GR, this expression is similar to one that appears in the derivation by Marklund and Clarkson \cite{10} in the context of GR dynamo MHD equation, where instead of the coupling between the magnetic vector potential and the Ricci tensor, one has the coupling of the Ricci tensor with the magnetic field itself. Actually in the next section one shall see that this last term shall appear in the equation of the mean field dynamo in Riemannian space with turbulent diffusion and kinetic helicity.\newpage

\section{Mean field dynamos in Riemannian space}
Though most of cosmological models do not may use of turbulent diffusivity, and in many situations the universe fluid can be considered as highly conductive \cite{11}, particle astrophysicists and cosmologists are faced with problems which goes from the cosmic magnetism origin to magnetism in the early universe \cite{11}. An alternative way of making computations easier shall be discussed in the next section when one shall compute the mean field dynamo idea to the Goedel idea, but before closing this section, and to exploit more physical cosmological implications, one shall consider the case of the electric field in terms of the Riemannian effects. This section considers a simple way to address the mean-field dynamos starting from the mean field equation \cite{4},
\begin{equation}
{\partial}_{t}\textbf{B}={\alpha}{\nabla}{\times}(\textbf{U}{\times}\textbf{B})+{\beta}{\Delta}\textbf{B}
\label{19}
\end{equation}
where the kinetic helicity is given by
\begin{equation}
{\alpha}=\frac{\tau}{3}\langle\textbf{U}.{\nabla}{\times}\textbf{U}\rangle=\langle\textbf{U}.{\omega}\rangle
\label{20}
\end{equation}
 and ${\beta}$ is the turbulent diffusion. As in Marklund-Clarkson derivation, the main effect of Riemann or Ricci curvature is coupled with diffusion. The only difference now, is that here the diffusion appears in the form of turbulent diffusion, and that to simplify matters the flow vectors are parallel transport along the curved Riemannian space. By using the Marklund-Clarkson computation of the Laplacian operator ${\Delta}={\nabla}^{2}$, yields \begin{equation}
 {\Delta}B^{i}=-curl(curl)B^{i}=-D^{2}B^{i}+D^{i}(D_{j}B^{j})+2{\epsilon}^{ijk}{\dot{B}}_{j}{\omega}_{k}+2R^{ij}B_{j}\label{21}
 \end{equation}
 By considered that all the vectors involved in the mean-field dynamos are parallel (Fermi) transport along the lines of the flow in 3D Riemannian curved spatial section of the cosmological model, thus
 \begin{equation}
 D_{j}B^{k}=0\label{22}
 \end{equation}
 which also implies $D^{2}B^{i}$ vanishes since $D^{2}=D_{i}D^{i}$ and $D^{i}$ is the covariant derivative operator. Thus the expression for the Laplacian of the magnetic field above reduces to
 \begin{equation}
 {\Delta}B^{i}=-curl(curl)B^{i}= 2\frac{1}{\sqrt{g}}{\epsilon}^{ijk}{\dot{B}}_{j}{\omega}_{k}+2R^{ij}B_{j}\label{21}
 \label{23}
 \end{equation}
 Substitution of this expression into the mean-field dynamo equation reads
 \begin{equation}
 {\gamma}(B^{i}- 2{\epsilon}^{ijk}{\dot{B}}_{j}{\omega}_{k})=2{\beta}R^{ij}B_{j}+{\alpha}[{\nabla}{\times}\textbf{B}]^{i}\label{21}
 \label{24}
 \end{equation}
 Thus by contracting this equation with the covariant components of the magnetic field $B_{i}$ and using Einstein summation rule, one obtains
 \begin{equation}
 {\gamma}(B^{2}- 2{\epsilon}^{ijk}{B}_{j}B_{i}{\omega}_{k})=2{\beta}R^{ij}B_{i}B_{j}+{\alpha}[\textbf{B}.{\nabla}{\times}\textbf{B}]\label{21}
 \label{25}
 \end{equation}
 Where one used here, the exponential growth rate ${\gamma}$ in the format $\textbf{B}=\textbf{B}_{0}(\textbf{r})e^{{\gamma}t}$. The last term inside the brackets is the current helicity ${\lambda}$. Due to the totally skew-symmetry property of the Levi-Civita symbol ${\epsilon}^{ijk}$, the second term on the LHS of the last equation vanishes. This reduces (\ref{25}) to
 \begin{equation}
 {\gamma}= 2{\beta}\frac{R^{ij}B_{i}B_{j}}{{B^{2}}}+{\alpha}{\lambda}
 \label{26}
 \end{equation}
 Note that in this case since \cite{12} $\beta$ is given by
 \begin{equation}
 \beta=\frac{vl}{3}\ge{0}
 \label{27}
 \end{equation}
 the growth rate ${\gamma}$ is positive, when both kinetic and current helicity possesses the same sign. Since ${\gamma}$ shall be positive even when the turbulent diffusivity vanishes, the dynamo action is fast \cite{14}. But the most important conclusion driven from the last mean-field dynamo expression is that turbulent diffusion couples with Riemann curvature, and in the case of the case of curved Riemannian space, the turbulent diffusivity effects would be enhanced by curvature. Thus one expects that in cosmological models, turbulence would be enhanced. Similar results had been obtained by Brandenburg et al \cite{13}. Other cosmological implications and consequences, shall be discussed in the next section.
 \newpage
 \section{Mean-field dynamos in cosmology}
 From expression (\ref{26}) an immeadiate cosmological implication is that the, if one only considers magnetic fields the Barrow-Tsagas expression \cite{11}
 \begin{equation}
 R_{ij}B^{i}B^{j}=\frac{1}{2}{\Pi}_{ij}B^{i}B^{j}+\frac{1}{3}B^{4}=0
 \label{28}
 \end{equation}
 and expression for the growth rate reduces to
 \begin{equation}
 {\gamma}= {\alpha}{\lambda}
 \label{29}
 \end{equation}
 Here ${\Pi}_{ij}$ is the anisotropic stress. In general this term of the contracted magnetic curvature in terms of the Ricci tensor $R_{ij}$ does not vanish. Actually the Ricci tensor is given by \cite{10}
 \begin{equation}
 {R^{ij}B_{i}B_{j}}=\frac{2}{3}[({\cal{E}}+{\Lambda}-\frac{1}{2}{\Theta}^{2}-{\sigma}^{2}+{\omega}^{2})h_{ij}+
 {\Pi}_{ij}-{\dot{\sigma}}_{ij}+D_{(i}{\dot{u}}_{j)}]B^{i}B^{j}
 \label{30}
 \end{equation}
 Here ${\Lambda}$ is the cosmological constant and ${\cal{E}}$ is the energy density, and ${\sigma}_{ij}$ represents the shear. Therefore, in the absence of shear, anisotropic stresses and rotation the Friedmann like universe mean-field dynamo growth rate would be given by
 \begin{equation}
 {\gamma}=\frac{2}{3}{\beta}({\cal{E}}+{\Lambda}-\frac{1}{2}{\Theta}^{2})\frac{h_{ij}B^{i}B^{j}}{B^{2}}
 \label{31}
 \end{equation}
 The kinetic helicity term vanishes since the kinetic helicity depends upon vorticity of the cosmological model which is absent in Friedmann cosmology. Since the three dimensional projective tensor $h_{ij}$ is given by
 \begin{equation}
 h_{ij}=g_{ij}+U_{i}U_{j}
 \label{32}
 \end{equation}
 Expression (\ref{31}) reduces to
 \begin{equation}
 {\gamma}=\frac{2}{3}{\beta}({\cal{E}}+{\Lambda}-\frac{1}{2}{\theta}^{2})[1+\frac{(\textbf{U}.\textbf{B})}{B^{2}}]
 \label{33}
 \end{equation}
 Since the last term inside the brackets is always positive, the growth rate shall give rise to a dynamo action if the term inside the round brackets is positive, but in general square of the expansion scalar ${\Theta}$ is strong which would imply a decaying of magnetic field, which has already been observed before \cite{11}. The same does not happen with the Goedel spacetime where the expansion rate vanishes and the vorticity appears in the mean-field growth rate as
\begin{equation}
{\gamma}=\frac{2}{3}{\beta}({\cal{E}}+{\Lambda}+{\omega}^{2})
\label{34}
\end{equation}
In this expression the later term inside the brackets dissapears since the frame used here is comoving and $\textbf{U}$ vanishes. This shall give rise to a slow mean-field dynamo cosmological model. Since in Goedel cosmology the cosmological constant may be considered to vanish and
\begin{equation}
{\cal{E}}={\omega}^{2}
\label{35}
\end{equation}
expression (\ref{34}) simplifies further to
\begin{equation}
{\gamma}=\frac{4}{3}{\beta}{\omega}^{2}\ge{0}
\label{36}
\end{equation}
Thus a dynamo action is granted in Goedel cosmology. Now let us investigate the astronomical consequences of this  mean-field cosmology. Thus the equation of the growth rate of the covariant mean field dynamo
\begin{equation}
{\gamma}= 2{\beta}\frac{\langle{\langle{Ric},\textbf{B}\rangle},\textbf{B}\rangle}{{B^{2}}}+{\alpha}{\lambda}
\label{37}
\end{equation}
yields the following growth rate in the radiation era \cite{14} where ${\omega}\approx{10^{-6}s^{-1}}$ as
\begin{equation}
{\gamma}=\frac{4}{3}{\beta}{\times}10^{-12}s^{-1}
\label{38}
\end{equation}
No dropping the comoving frame and considering that $\textbf{u}.\textbf{B}=1$ and that $B^{2}$ is very small, thus
\begin{equation}
B=\sqrt{2\frac{\beta}{\gamma}}{\omega}
\label{39}
\end{equation}
which shows that by considering the Ruzmaikin-Sokoloff \cite{2} result $B_{0}\ge{{\omega}{\times}10^{-10}G}$, the last expression yields $B>10^{-24}G$ which is certainly well-within the $B\approx{10^{-21}G}$ limit which is necessary to seed the galactic dynamos. This shows not only, that the Riemannian mean-field dynamos are well within the galactic dynamos, but also that it serves as a good model for cosmological dynamos.
\newpage
\section{The spectrum of Ricci scalar in MHD GR dynamo turbulence}
In this section the de Sitter metric
\begin{equation}
ds^{2}=dt^{2}-exp{{\Lambda}t}(dx^{2}+dy^{2}+dz^{2})
\label{40}
\end{equation}
and its Ricci scalar $R(x)=6{\Lambda}$, where ${\Lambda}$ is the cosmological constant are used to compute the energy spectrum of the turbulent dynamo. This can be done by
considering the Fourier spectrum of the curvature scalar and the above equation for GR-MHD dynamo equation. To achieve this aim, let us first Fourier analyze the Marklund-Clarkson equation GR-MHD dynamo equation according to the rule
\begin{equation}
B(\textbf{x},t)=exp{[i\textbf{k.x}+{\gamma}(k)t]}b_{0}(\textbf{k})
\label{41}
\end{equation}
where ${\gamma}(k)$ is the Fourier transform of the growth rate of magnetic fields. The factor k is the dynamo wave number. Thus the GR-MHD dynamo equation in the spectrum format is
\begin{equation}
{\gamma}(k,{\eta})=-\frac{1}{3}{\eta}\frac{[R(k)+k^{2}]}{[1+\frac{4}{3}{\Theta}{\eta}]}
\label{42}
\end{equation}
For long wavelength ${\lambda}=\frac{2\pi}{k}$ or short wave number $k$, and considering that expansion ${\Theta}=div\textbf{v}=ikv_{0}$ where $v_{0}$ is a stationary velocity, one obtains that
\begin{equation}
\textbf{Re}{\gamma}(k,{\eta})= -\frac{1}{3}{\eta}{[R(k)+k^{2}]}
\label{43}
\end{equation}
This implies that the limit when ${\eta}\rightarrow{0}$ is
\begin{equation}
lim_{{\eta}\rightarrow{0}}\textbf{Re}{\gamma}(k,{\eta})=0
\label{44}
\end{equation}
which shows that the dynamo is slow. Fourier analyzing the above formula for the Ricci scalar it can be expressed in terms of the growth rate factor ${\gamma}$ as
\begin{equation}
R(k)=-{\Lambda}k^{-1}
\label{45}
\end{equation}
The curvature spectrum density defined by
\begin{equation}
R^{2}(k)={\Lambda}^{2}k^{-2}
\label{46}
\end{equation}
is shown to lead to the following energy turbulent spectrum
\begin{equation}
E_{Ric}(k)={\Lambda}^{2}k^{-2}
\label{47}
\end{equation}
This formula shows that in the short wavelength limit, the energy spectrum of the turbulent dynamo is similar to  Kolmogorov one
\begin{equation}
E(k)\approx{k^{-\frac{5}{3}}}\approx{k^{-1.88}}
\label{47}
\end{equation}
Thus one must conclude that the energy spectrum of the turbulent dynamo in de Sitter universe is similar to the Kolmogorov one. Note that when $k<<1$, the curvature energy and Ricci scalar grow with dynamo wavenumber.
Note that since the cosmological constant is very small the turbulent Ricci energy is also damped by the cosmic constant. \section{Conclusions}
A previous model of cosmic dynamo, has been proposed by Bassett et al \cite{15} by making use of pre-heating phases of inflationary models. Other interesting model for cosmic magnetism in terms of dynamos has been proposed also by Enqvist \cite{16} who proposed to used small scale random magnetic fields to explain galactic magnetic fields. Nevertheless the author has not considered the mean-field dynamo equation. Here an explicitly use is made of the mean-fields in Riemannian space as the three-dimensional cosmological section of Friedmann and Goedel universes. Similar to the magnetic field obtained by Enqvist of $B\approx{10^{-20}}$, one here obtains a magnetic field lower bound of $10^{-24}G$, which includes also the $10^{-21}G$ usually found in the literature \cite{15}. Here one has considered the isotropic case where ${\alpha}_{ij}={\alpha}g_{ij}$ where ${\alpha}_{ij}$ is the kinetic helicity tensor and $g_{ij}$ is the Riemannian tensor. Future prospects include the most complicated issue to find out a general mean field dynamo theory in Riemannian curved space or the pseudo-Riemannian spacetime of Einstein general relativity. Fast mean field dynamos \cite{17} has been obtained in Goedel model. Recently Chicone et al \cite{18} has also considered a fast dynamo in two-dimensional compact Riemannian space that does not violate Cowling's anti-dynamo theorem, and this can provide also a nice model to deal with some cosmological models that possesses Riemannian 3D spatial sections such as the Friedmann hyperbolic spaces. By analyzing the Fourier modes of the Ricci scalar. one shows that the energy spectrum of the turbulent dynamo is similar to the Kolmogorov spectrum. By considering a universe vorticity, ${\omega}\approx{10^{-16}s^{-1}}$ for galactic dynamos, thus ${\gamma}=2.10^{-32}{\beta}$, and since ${\beta}\approx{10^{26}cm^{2}s^{-1}}$, the growth rate ${\gamma}\approx{10^{-6}s^{-1}}$. In non-comoving the magnetic field is expressed as $B\approx{\sqrt{\frac{2{\beta}}{\gamma}}{\times}10^{-6}G}\approx{10^{10}G}$ a magnetic field found in the nucleosynthesis era. Recently Mizena, Stepanov and Frick \cite{19} have shown that the energy spectrum above can grow from ${1.88}$ to $2$, in its exponents, when cross-helicity is present. Some of the results in this paper have been reproduced by Bassett et al \cite{15} without recurring to the mean field dynamos in the case of Friedmann-Robertson-Walker cosmological model.

\newpage
\section{Acknowledgements}
Several discussions with D Sokoloff, K Raedler and R Beck on possibilities of mean field dynamo theory in Riemannian space, are highly appreciated. I also thank financial supports from UERJ and CNPq.
 
  \end{document}